\documentclass[12pt]{iopart}
\usepackage{graphicx,,subfigure}

\begin{document}

\title[A new scheme for probabilistic teleportation and its potential applications]{A new scheme for probabilistic teleportation and its potential applications}

\author{Jiahua Wei$^1$, Hong-Yi Dai$^2$, Ming Zhang$^1$${^{,\dag}}$}

\address{$^1$Department of Automatic Control, College of Mechatronics and Automation, National University of Defense Technology, Changsha, Hunan, 410073, People's Republic of China

$^2$Department of Physics, College of Science, National University of Defense Technology, Changsha, Hunan, 410073, People's Republic of China}
\ead{${^\dag}$zhangming@nudt.edu.cn, weijiahua@126.com}

\begin{abstract}
We propose a novel scheme to probabilistically teleport an unknown two-level quantum state when the information of the partially entangled state is only available for the sender. This is in contrast with the fact that the receiver must know the non-maximally entangled state in previous typical schemes for the teleportation. Additionally, we illustrate two potential applications of the novel scheme for probabilistic teleportation from a sender to a receiver with the help of an assistant, who plays distinct roles under different communication conditions, and our results show that the novel proposal could improve the security and enlarge the applied range of probabilistic teleportation.
\end{abstract}

\section{Introduction}
Quantum teleportation is the process that transmits quantum states from a sender to a remote receiver via a quantum channel with the help of some classical information. It was first proposed by Bennett et al.\cite{Bennett} in 1993, and experimentally demonstrated by Bouwmeester et al.\cite{Bouwmeester} in 1997. Since then, the development of quantum teleportation has gone through twenty years, and some theoretical and experimental progresses\cite{Boschi,Karlsson,Li,Koniorczyk,Werner,Dai1,Dai2,Yeo,Zhang2,Dai4,Tanaka,Neves,Zhang6,Chiribella} have been made in this domain.

For the sake of that quantum teleportation could be applied to quantum computation and quantum information\cite{Nielsen}, a growing number of works have appeared in the teleportation recently. Li et al.\cite{Li} proposed a scheme of probabilistic teleportation to transmit an unknown single-qubit using a non-maximally entangled state when the receiver introduced an auxiliary particle. Dai et al.\cite{Dai1,Dai2} presented two protocols for probabilistically teleporting an arbitrary two-particle state from a sender to either one of two receivers via two partially entangled W states and by the combination of a non-maximally entangled GHZ state and an entangled W state, respectively. Meanwhile, some authors have generalized these methods to the case of the higher-dimensional quantum state\cite{Dai3,Peng,Ikram,Evers}. In Ref. \cite{Evers}, a protocol for the teleportation of arbitrary quantum states of four-dimensional qudits is presented. Nevertheless, there are also many important and open subjects to be taken into account for quantum teleportation.

In most schemes about probabilistic teleportation of quantum states using a partially entangled state as a quantum channel, the receiver Bob not only must introduce an auxiliary particle and make a corresponding unitary transformation to reconstruct the original quantum state, but also needs to fully know the information of the non-maximally entangled state. Evidently, the previous schemes are not valid on condition that only the sender Alice has total knowledge of the partially entangled state. In order to overcome this drawback, we proposed a novel protocol to probabilistically teleport an unknown quantum state to the receiver Bob when the information of the partially entangled state is only available for the sender Alice, i.e., the receiver Bob does not know the information of the non-maximally entangled state completely. Then a comparison has been made between our present protocol and the former ones.

Furthermore, we illustrate two concrete applications of our scheme for probabilistic teleportation of an unknown two-level quantum state from the sender Alice to the receiver Bob with the aid of the assistant Charlie under different communication conditions. Assuming that the sender Alice wants to teleport an unknown two-level quantum state to the receiver Bob with a restriction that there is no quantum channel between the sender Alice and the receiver Bob. Fortunately, the assistant Charlie has two quantum channels with the sender Alice and the receiver Bob, respectively. Thus the sender Alice and the receiver Bob could accomplish the teleportation task in spite of the fact that the two quantum channels are only available for the assistant Charlie, who plays a leading role in the whole probabilistic teleportaton from the sender Alice to the receiver Bob. On the other hand, when only the sender Alice and the receiver Bob know the two quantum channels, and the assistant Charlie just plays a role of repeater between the sender Alice and the receiver Bob, the teleportation task is still realizable according to our scheme. Hence the roles of the assistant Charlie under different communication conditions are quite distinct. It should be emphasized that the novel proposal is essential to perform the teleportation processes under different communication conditions, and this proposal could improve the security of the teleportation processes and broaden the applied range of probabilistic teleportation.

The rest of this paper are organized as follows: A previous typical scheme of probabilistic teleportation is stated in Section 2, and the correlative unitary transformations for the receiver are given in detail. In Section 3, we propose a novel scheme for probabilistically transmitting an unknown quantum state via generalized measurements when the information of the partially entangled state is only available for the sender. The concrete implementation processes of this scheme are elaborated. Moreover, the main differences of our proposal from the former ones are presented in this section. Two concrete applications of the new scheme are given to indicate the advantage of our proposal in Section 4. The paper concludes with Section 5.

\section{A typical scheme of probabilistic teleportation}
Let us begin with a brief statement of probabilistic teleportation\cite{Li} using a partially entangled state. Suppose that the sender Alice wants to transmit the following unknown quantum state to the receiver Bob separated spatially
\begin{equation}\label{1}
    |\psi_1\rangle=\alpha|0_1\rangle+\beta|1_1\rangle
\end{equation}
where $\alpha$ is real and $\beta$ is a complex number, and $|\alpha|^2+|\beta|^2=1$. Without loss of generality, the entanglement channel is composed of a partially entangled two-particle state below
\begin{equation}\label{2}
    |\psi_{23}\rangle=a|0_20_3\rangle+b|1_21_3\rangle
\end{equation}
where the real coefficient $a$ and the complex one $b$ satisfy $|a|^2+|b|^2=1$, and $|a|\geq|b|>0$. Particle 2 belongs to the sender Alice, while particle 3 belongs to the receiver Bob. Therefore, the state of the whole system composed of particles 1, 2 and 3 is given by
\begin{eqnarray}\label{3}
  |\psi_{total}\rangle &=& |\psi_1\rangle\otimes|\psi_{23}\rangle \nonumber\\
   & =&\alpha{a}|0_10_20_3\rangle+\alpha{b}|0_11_21_3\rangle+\beta{a}|1_10_20_3\rangle+\beta{b}|1_11_21_3\rangle
\end{eqnarray}

In order to realize the teleportation, Alice needs to perform the following Bell-state measurements on particles 1 and 2.
\begin{eqnarray}\label{4-5}
  |\phi^\pm_{12}\rangle &=& \frac{1}{\sqrt{2}}(|0_10_2\rangle\pm|1_11_2\rangle) \\
  |\Psi^\pm_{12}\rangle &=& \frac{1}{\sqrt{2}}(|0_11_2\rangle\pm|1_10_2\rangle)
\end{eqnarray}
Then the total state will collapse into one of the four kinds of outcomes below
\begin{eqnarray}\label{6-7}
\langle\phi^\pm_{12}|\psi_{total}\rangle &=& \frac{1}{\sqrt{2}}(\alpha{a}|0_{3}\rangle\pm\beta{b}|1_{3}\rangle) \\
\langle\Psi^\pm_{12}|\psi_{total}\rangle &=& \frac{1}{\sqrt{2}}(\alpha{b}|1_{3}\rangle\pm\beta{a}|0_{3}\rangle)
\end{eqnarray}

After the Bell-state measurements, Alice informs Bob of her measurement results via a classical channel. For the purpose of obtaining the original state shown in Eq. (1), Bob needs to introduce an auxiliary particle \emph{m} with an initial state $|0_{m}\rangle$, and performs a conditional unitary transformation $U_F$ on particles 3 and \emph{m}, which depends on the Bell-state measurement results. Table 1 shows the results after the measurements on particles 1, 2 and the unitary transformation on particles 3 and \emph{m}.
\begin{table}[!htb]
  \centering
  \caption{The Bell-state measurement results(BMRs) and the transformation $U_F$}
  \label{Table 1}
  \begin{tabular*}{\textwidth}{c|c|c|c|c}
    \hline
    \hline
     BMRs  on &  & \multicolumn{3}{c}{Results after the transformation $U_F$}\\
     \cline{3-5}
    particles 1, 2& \raisebox{1.5ex}[0pt]{~~~~~~$U_F$~~~~~~} & States of particle \emph{m} & States of particle 3 & Probabilities \\ \hline \hline
     &  & $|0_{m}\rangle$ & $\alpha|0_3\rangle+\beta|1_3\rangle$ & $\frac{|b|^2}{2}$\\
    \cline{3-5}
     \raisebox{1.5ex}[0pt]{$|\phi^+_{12}\rangle$}&  \raisebox{1.5ex}[0pt]{$U_F^0$} & $|1_{m}\rangle$ & $|0_3\rangle$ & $\frac{|\alpha|^2}{2}(1-2|{b}|^2)$ \\
     \cline{1-2} \cline{3-5}
     &  & $|0_{m}\rangle$ & $\alpha|0_3\rangle+\beta|1_3\rangle$ & $\frac{|b|^2}{2}$\\
    \cline{3-5}
     \raisebox{1.5ex}[0pt]{$|\phi^-_{12}\rangle$} & \raisebox{1.5ex}[0pt]{$U_F^1$} & $|1_{m}\rangle$ & $|0_3\rangle$ & $\frac{|\alpha|^2}{2}(1-2|{b}|^2)$ \\
     \cline{1-2} \cline{3-5}
     &  & $|0_{m}\rangle$ & $\alpha|0_3\rangle+\beta|1_3\rangle$ & $\frac{|b|^2}{2}$\\
    \cline{3-5}
    \raisebox{1.5ex}[0pt]{$|\Psi^+_{12}\rangle$} & \raisebox{1.5ex}[0pt]{$U_F^2$} & $|1_{m}\rangle$ & $|1_3\rangle$ & $\frac{|\beta|^2}{2}(1-2|{b}|^2)$ \\
     \cline{1-2} \cline{3-5}
     &  & $|0_{m}\rangle$ & $\alpha|0_3\rangle+\beta|1_3\rangle$ & $\frac{|b|^2}{2}$\\
    \cline{3-5}
    \raisebox{1.5ex}[0pt]{$|\Psi^-_{12}\rangle$} & \raisebox{1.5ex}[0pt]{$U_F^3$} & $|1_{m}\rangle$ & $|1_3\rangle$ & $\frac{|\beta|^2}{2}(1-2|{b}|^2)$ \\
    \cline{1-2}  \cline{3-5}
    \hline
    \hline
  \end{tabular*}
\end{table}

The unitary transformations $U^i_F(i=0,1,2,3)$ in Table 1 are described as
\begin{equation}\label{8}
    \begin{array}{cc}
    U^0_F=\left(
       \begin{array}{cc}
         A(a,b) & \textbf{0} \\
         \textbf{0} & \sigma_z \\
       \end{array}
     \right)
      &
     U^1_F=\left(
        \begin{array}{cc}
          A(a,b) & \textbf{0} \\
          \textbf{0} & -\sigma_z \\
          \end{array}
        \right)
        \\
     U^2_F=\left(
        \begin{array}{cc}
          \textbf{0} & \sigma_z \\
          A(a,b) & \textbf{0} \\
          \end{array}
        \right)
       &
      U^3_F=\left(
        \begin{array}{cc}
          \textbf{0} & -\sigma_z \\
          A(a,b) & \textbf{0} \\
          \end{array}
        \right)
     \end{array}
\end{equation}
\noindent where $\textbf{0}$ is the $2\times2$ zero matrix, $\sigma_z$ is known as one of pauli matrixes, and $A(a,b)$ is the $2\times2$ matrix relative to the parameters $a$ and $b$ of Eq. (2). $\sigma_z$ and $A(a,b)$ could be expressed as
\begin{eqnarray}\label{9}
         \sigma_z=\left(
       \begin{array}{cc}
         1 & 0 \\
         0 & -1
       \end{array}
        \right)
     &\quad &
       A(a,b)=\left(
       \begin{array}{cc}
         \frac{b}{a} & \sqrt{1-\frac{|b|^2}{|a|^2}} \\
         \sqrt{1-\frac{|b|^2}{|a|^2}} & -\frac{b}{a}
       \end{array}
        \right)
\end{eqnarray}

Subsequently, Bob measures the sate of particle \emph{m}. If the result is $|1_{m}\rangle$, quantum teleportation fails. Otherwise, when the result is $|0_{m}\rangle$, the teleportation will be realized with the successful probability of $\frac{|b|^2}{2}$. It is demonstrated that quantum teleportation is successfully obtained with the equal probability under the four kinds of Bell-state measurement results\cite{Li}, hence the whole successful probability of quantum teleportation is $2|b|^2$. If $|a|=|b|=\frac{1}{\sqrt{2}}$, i.e., the quantum channel is composed of a maximally entangled state, the total probability is equal to one.

\section{The teleportation when only the sender knows the entangled state}
We would like to point out that the receiver Bob must know the parameters $a$ and $b$ shown in Eq. (2) of the partially entangled state to perform the relevant unitary transformation $U_F$ on particles 3 and \emph{m} in the previous typical schemes\cite{Li,Dai1,Dai2}. It is revealed that one can not make use of this scheme to realize probabilistic teleportation in the situation that only the sender Alice has full knowledge of the non-maximally entangled state in the teleportation processes. Based on this observation, we propose a novel scheme to transmit an unknown state, whether the receiver knows the partially entangled state or not. Moreover, the detailed processes of our proposal are elaborated in this section.

Based on Eq. (3), the whole system could be expressed as follows
\begin{eqnarray}\label{10}
  |\psi_{total}\rangle &=&\frac{1}{2}(a|0_10_2\rangle+b|1_11_2\rangle)(\alpha|0_3\rangle+\beta|1_3\rangle) \nonumber\\
   &+& \frac{1}{2}(a|0_10_2\rangle-b|1_11_2\rangle)(\alpha|0_3\rangle-\beta|1_3\rangle) \nonumber\\
   &+& \frac{1}{2}(b|0_11_2\rangle+a|1_10_2\rangle)(\alpha|1_3\rangle+\beta|0_3\rangle) \nonumber\\
   &+& \frac{1}{2}(b|0_11_2\rangle-a|1_10_2\rangle)(\alpha|1_3\rangle-\beta|0_3\rangle)
\end{eqnarray}

Actually, our scheme needs to take advantage of a generalized measurement\cite{Nielsen,Zhang} with the five measurement operators ${M_i (i=0,1,2,3,4)}$.
\begin{eqnarray}\label{11}
  M_0 &=& \frac{1}{\sqrt{2}|a|}(a|0_10_2\rangle+b|1_11_2\rangle)(b\langle0_10_2|+a\langle1_11_2|) \nonumber\\
  M_1 &=& \frac{1}{\sqrt{2}|a|}(a|0_10_2\rangle-b|1_11_2\rangle)(b\langle0_10_2|-a\langle1_11_2|) \nonumber\\
  M_2 &=& \frac{1}{\sqrt{2}|a|}(b|0_11_2\rangle+a|1_10_2\rangle)(a\langle0_11_2|+b\langle1_10_2|) \nonumber\\
  M_3 &=& \frac{1}{\sqrt{2}|a|}(b|0_11_2\rangle-a|1_10_2\rangle)(a\langle0_11_2|-b\langle1_10_2|) \nonumber\\
  M_4 &=& \sqrt{1-\frac{|b|^2}{|a|^2}}({|0_10_2\rangle}\langle0_10_2|+{|1_10_2\rangle}\langle1_10_2|)
\end{eqnarray}
It is underlined that the aforementioned operators satisfy the completeness relation $\sum_{i=0}^{4}M_i^\dag{M_i}=I$. Afterwards, Alice performs the generalized measurements on particles 1 and 2, and informs Bob of the measurement results with the help of some classical information. In order to reconstruct the original state, Bob performs a relevant unitary transformation $U_G$ on particles 3. Table 2 shows the results after the generalized measurements on particles 1, 2 and the unitary transformation $U_G$ on particle 3. Besides, the total successful probability of quantum teleportation is also equal to $2|b|^2$. When the quantum channel is composed of a maximally entangled state, the total successful probability equals one.
\begin{table}
  \centering
  \caption{The generalized measurement results(GMRs) and the transformation $U_G$}\label{Table 2}
  \begin{tabular*}{\textwidth}{c|c|c|c}
    \hline
    \hline
     &  & States of particle 3  &   \\
      \raisebox{1.5ex}[0pt]{Measurement operators} & \raisebox{1.5ex}[0pt]{~~Probabilities~~} & ~~after the GMRs~~ & \raisebox{1.5ex}[0pt]{~~~~~~~~$U_G$~~~~~~}  \\
     \hline
     \hline
    $M_0$ & $\frac{|b|^2}{2}$ & $\alpha|0_3\rangle+\beta|1_3\rangle$ & $I$  \\ \hline
    $M_1$ & $\frac{|b|^2}{2}$ & $\alpha|0_3\rangle-\beta|1_3\rangle$ & $\sigma_z$  \\ \hline
    $M_2$ & $\frac{|b|^2}{2}$ & $\alpha|1_3\rangle+\beta|0_3\rangle$ & $\sigma_x$  \\ \hline
    $M_3$ & $\frac{|b|^2}{2}$ & $\alpha|1_3\rangle-\beta|0_3\rangle$ & $i\sigma_y$  \\ \hline
    $M_4$ & $1-2|b|^2$ & $|0\rangle_3$ &$--$\\ \hline
    \hline
  \end{tabular*}
\end{table}

In reality, projective measurements together with unitary operations are sufficient to implement a generalized measurements with the aid of the introduction of auxiliary particles\cite{Nielsen}. The concrete implementation procedures of this new scheme are presented as follows:

\textbf{\emph{Step 1}}: Alice introduces an auxiliary particle \emph{m} with an initial state $|0_{m}\rangle$, then the system could be written as
\begin{eqnarray}\label{12}
  |\psi^0_{12m3}\rangle &=& |\psi_{12}\rangle\otimes|0_m\rangle\otimes|\psi_{3}\rangle \nonumber\\
   &=&\alpha{a}|0_10_20_m0_3\rangle +\alpha{b}|0_11_20_m1_3\rangle \nonumber\\
   &+&\beta{a}|1_10_20_m0_3\rangle+\beta{b}|1_11_20_m1_3\rangle
\end{eqnarray}

\textbf{\emph{Step 2}}: On particles 1, 2 and \emph{m}, Alice performs an unitary transformation $U_S$, which takes the form of the following $8\times8$ matrix
\begin{equation}\label{13}
    U_S=\left(
       \begin{array}{cc}
         U_F^0 & \textbf{0}\\
         \textbf{0} & U_F^0
       \end{array}
     \right)
\end{equation}
where $\textbf{0}$ is the $4\times4$ zero matrix, and $U_F^0$ is given by Eq. (8). Then we can express the system as
\begin{eqnarray}\label{14}
    &&|\psi^1_{12m3}\rangle=(U_S|\psi^0_{12m}\rangle)\otimes|\psi^0_{3}\rangle \nonumber\\
    &&=\frac{b}{\sqrt{2}}\cdot\frac{1}{\sqrt{2}}(|0_10_2\rangle+|1_11_2\rangle)\otimes|0_{m}\rangle\otimes(\alpha|0_3\rangle+\beta|1_3\rangle)\nonumber\\
    &&+\frac{b}{\sqrt{2}}\cdot\frac{1}{\sqrt{2}}(|0_10_2\rangle-|1_11_2\rangle)\otimes|0_{m}\rangle\otimes(\alpha|0_3\rangle-\beta|1_3\rangle)\nonumber\\
    &&+\frac{b}{\sqrt{2}}\cdot\frac{1}{\sqrt{2}}(|0_11_2\rangle+|1_10_2\rangle)\otimes|0_{m}\rangle\otimes(\alpha|1_3\rangle+\beta|0_3\rangle)\nonumber\\
    &&+\frac{b}{\sqrt{2}}\cdot\frac{1}{\sqrt{2}}(|0_11_2\rangle-|1_10_2\rangle)\otimes|0_{m}\rangle\otimes(\alpha|1_3\rangle-\beta|0_3\rangle)\nonumber\\
    &&+{a}\sqrt{1-\frac{|b|^2}{|a|^2}}(\alpha|0_10_2\rangle+\beta|1_10_2\rangle)\otimes|1_{m}\rangle\otimes|0_3\rangle
\end{eqnarray}

\textbf{\emph{Step 3}}: Alice measures the state of this auxiliary particle \emph{m}, and performs the Bell-state measurements on particles 1 and 2. Subsequently, Alice informs Bob of her measurement results using a classical channel. It should be emphasized that the teleportation could be realized successfully when the state of particle \emph{m} is $|0_m\rangle$ with the probability of $2|b|^2$, otherwise it fails.

\textbf{\emph{Step 4}}: Based on these measurement results, Bob only needs to perform a relevant unitary transformation $U_T$ on particle 3 to obtain the original state. Table 3 indicates how to select the unitary transformation for particle 3 based on Alice's measurements results.
\begin{table}
  \centering
  \caption{The unitary transformation $U_T$ based on Alice's measurements results}
  \label{Table 3}
  \begin{tabular*}{\textwidth}{c|c|c|c}
    \hline
    \hline
    \multicolumn{2}{c|}{Measurement results}&  &   \\
    \cline{1-2}
    ~~States of particle \emph{m}~~ &~~States of particles 1, 2~~ & \raisebox{1.5ex}[0pt]{~~~Probabilities~~~} & \raisebox{1.5ex}[0pt]{~~~~~~$U_T$~~~~~~}  \\
     \hline
     \hline
     & $|\phi^+_{12}\rangle$ & $\frac{|b|^2}{2}$ & $I$  \\
      \cline{2-4}
     & $|\phi^-_{12}\rangle$ & $\frac{|b|^2}{2}$ & $\sigma_z$  \\
     \cline{2-4}
     \raisebox{1.5ex}[0pt]{$|0_{m}\rangle$}& $|\Psi^+_{12}\rangle$ & $\frac{|b|^2}{2}$ & $\sigma_x$  \\
     \cline{2-4}
     & $|\Psi^-_{12}\rangle$ & $\frac{|b|^2}{2}$ & $i\sigma_y$  \\
     \hline
     {$|1_{m}\rangle$}& $--$ & $1-2{|b|^2}$ & $--$  \\
    \hline
    \hline
  \end{tabular*}
\end{table}

\textbf{\emph{Remarks}}: According to the aforementioned analyses, we would like to point out that:
\begin{enumerate}
\item It should be underlined that one can use our scheme to carry out probabilistic teleportation when only the sender Alice fully knows the partially entangled state. This is different from the former typical one that the receiver Bob must have complete information of the entangled state.
\item Compared with the previous scheme\cite{Li} for probabilistic teleportation, the sender Alice in our current proposal needs an auxiliary particle, while one additional particle must be introduced by the receiver Bob in the former proposal.
\item The cost of an entangled state is  necessary for both the previous  typical scheme and our new scheme. In addition, classical communication is also essential to realize probabilistic teleportation.
\item The total successful probability for the teleportation of our presnet scheme is the same as the former one, and equals to $2|b|^2$, and the successful probability would be equal to one when the quantum channel is composed of a maximally entangled state.
\end{enumerate}

\section{Two potential applications of the novel scheme}
To illustrate that our proposal could be used to improve the security and extend the applied range of quantum teleportation, we will present two concrete applications of our scheme for probabilistically transmitting an unknown quantum state with the help of an assistant under different communication conditions.

Supposing that the sender Alice wants to transmit an unknown quantum state given by Eq. (1) to the receiver Bob, and there is no quantum channel between the sender Alice and the receiver Bob. Fortunately, the assistant Charlie has two quantum channels with the sender Alice and the receiver Bob, respectively. Without loss of generality, the two entanglement channels are expressed as follows
\begin{eqnarray}\label{15-16}
  && |\psi_{A{C_1}}\rangle=c|0_A0_{C_1}\rangle+d|1_A1_{C_1}\rangle \\
  && |\psi_{{C_2}B}\rangle=e|0_{C_2}0_B\rangle+f|1_{C_2}1_B\rangle
\end{eqnarray}
where the real coefficients $c$, $e$ and the complex ones $d$, $f$ satisfy $|c|^2+|d|^2=1$ ($|c|\geq|d|>0$) and $|e|^2+|f|^2=1$ ($|e|\geq|f|>0$). Particle A belongs to the sender Alice, and particle B belongs to the receiver Bob, while particles $C_1$ and $C_2$ belong to the assistant Charlie. As a matter of fact, the implementation processes of this total probabilistic teleportation task from the sender Alice to the receiver Bob could be reduced to two parties: at first Alice could teleport the original state to Charlie, and then Charlie transmits this state to Bob. Based on the two different communication conditions, we will take advantage of our proposal to realize the total probabilistic teleportation with the help of the assistant.

\subsection{When only Charlie knows the two partially entangled states $|\psi_{A{C_1}}\rangle$ and $|\psi_{{C_2}B}\rangle$}
If only the assistant Charlie has full information about the two non-maximally entangled states $|\psi_{A{C_1}}\rangle$ and $|\psi_{{C_2}B}\rangle$, the following concrete parts reveal that the assistant plays a leading role in the whole probabilistic teleportaton from the sender Alice to the receiver Bob. Figure 1 shows the schematic diagram of the total probabilistic teleportaton on the condition that the two partially entangled states $|\psi_{A{C_1}}\rangle$ and $|\psi_{{C_2}B}\rangle$ are only available for the assistant Charlie.

\begin{figure}\label{figure.1}
\centering
\scalebox{0.66}{\includegraphics{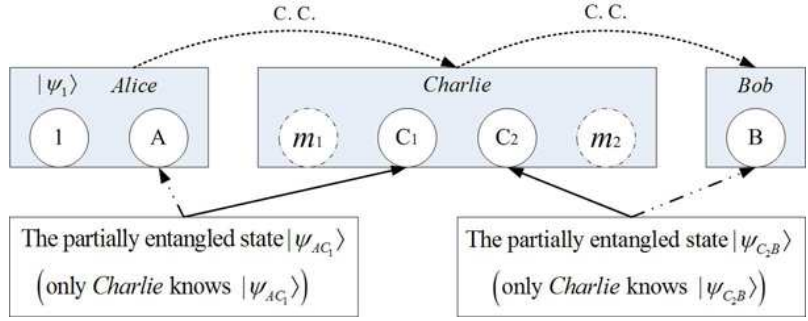}} 
\vspace*{1pt}
\caption{A sketch of the whole probabilistic teleportaton processes on the condition that the two partially entangled states $|\psi_{A{C_1}}\rangle$ and $|\psi_{{C_2}B}\rangle$ are only available for the assistant Charlie. Particle A belongs to the sender Alice, and particle B belongs to the receiver Bob, while particles $C_1$ and $C_2$ belong to the assistant Charlie. The state $|\psi_1\rangle$ of particle 1 is unknown. First, Alice uses the typical scheme to probabilistically transmit the original state to Charlie, who must introduce an auxiliary particles $m_1$. Subsequently, Charlie should introduce an auxiliary particles $m_2$ and utilize our novel proposal for probabilistically teleporting this state to Bob. C.C. represents classical communication.}
\end{figure}

\textbf{\emph{Part 1}}: Alice makes use of the typical scheme to probabilistically transmit the unknown quantum state to Charlie, who could be treated as the receiver in this part. To our knowledge, Charlie knows the non-maximally entangled state $|\psi_{A{C_1}}\rangle$, therefore the typical scheme is valid. According to Section 2, we can get that the teleportation from Alice to Charlie would be realized with the successful probability of $2|d|^2$. If this teleportation is successful, the original quantum state shown in Eq. (1) is reconstructed on particle $C_1$.

\textbf{\emph{Part 2}}: Charlie utilizes our novel proposal for probabilistically teleporting the state of particle $C_1$ to Bob. It should be pointed out that the receiver Bob has no knowledge about the non-maximally state $|\psi_{{C_2}B}\rangle$, hence the previous typical scheme can not be applied in this situation, and one can adopt the novel proposal for probabilistic teleportation from Charlie to Bob. Section 3 indicates that the state of particle $C_1$ should be reconstructed on particle $B$ with the successful probability of $2|f|^2$.

In the total probabilistic teleportation processes, the sender Alice and the receiver Bob accomplish the teleportation task with the successful probability of $4|df|^2$ in spite of the fact that only the assistant Charlie completely knows the partially entangled states $|\psi_{A{C_1}}\rangle$ and $|\psi_{{C_2}B}\rangle$. Besides, both of the sender Alice and the receiver Bob do not know whether the total probabilistic teleportation successes or not, and only the assistant Charlie knows the answer to this question, thus the security of the total probabilistic teleportation could be enhanced.

\subsection{When only Alice knows the partially entangled state $|\psi_{A{C_1}}\rangle$, and only Bob knows the partially entangled state $|\psi_{B{C_2}}\rangle$}
In this subsection, we will explore the teleportation problem when the assistant Charlie does not have any information about the two partially entangled states $|\psi_{A{C_1}}\rangle$ and $|\psi_{{C_2}B}\rangle$, and propose another potential application of the novel scheme. In this application, the assistant Charlie plays a role of a passive repeater between the sender Alice and the receiver Bob in the whole probabilistic teleportaton. Figure 2 shows the schematic diagram of the total probabilistic teleportaton when only Alice knows the partially entangled state $|\psi_{A{C_1}}\rangle$, and only Bob knows the partially entangled state $|\psi_{B{C_2}}\rangle$.

\begin{figure}\label{figure.2}
\centering
\scalebox{0.66}{\includegraphics{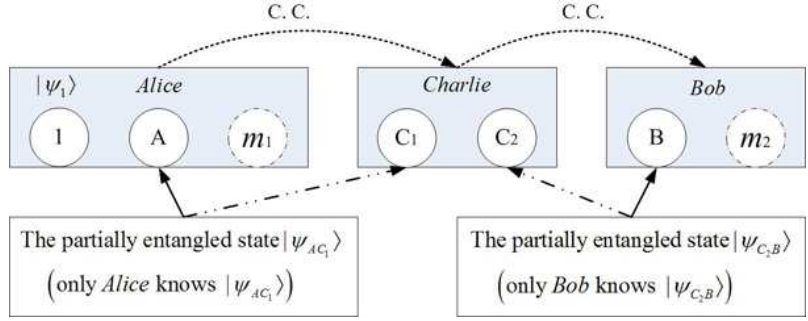}} 
\vspace*{1pt}
\caption{A sketch of the whole probabilistic teleportaton processes when only the sender Alice knows the partially entangled state $|\psi_{A{C_1}}\rangle$, and only the receiver Bob knows the partially entangled state $|\psi_{B{C_2}}\rangle$, but the assistant Charlie does not have any information about the two partially entangled states. First, Alice needs to introduce an auxiliary particles $m_1$ and utilize our novel scheme to probabilistically transmit the original state to Charlie. Then Charlie would use the typical proposal for probabilistically teleporting this state to Bob, who should introduce an auxiliary particles $m_2$.}
\end{figure}

\textbf{\emph{Part 1}}: Alice takes advantage of our scheme to probabilistically transmit the unknown quantum state to Charlie. It should be emphasized that the assistant Charlie, as the receiver in this part, does not know the non-maximally state $|\psi_{A{C_1}}\rangle$, and the novel proposal is the only choice for probabilistic teleportation from Alice to Charlie. Besides, the original state is reconstructed on particle $C_1$ with the successful probability of $2|d|^2$.

\textbf{\emph{Part 2}}: Charlie uses the typical proposal for probabilistically teleporting the state of particle $C_1$ to Bob. As a result of that the partially entangled state $|\psi_{{C_2}B}\rangle$ is available for Bob, the typical scheme is useful, and the probabilistic teleportation from Charlie to Bob would be performed with the successful probability of $2|f|^2$.

Even though the assistant Charlie does not have information of the two quantum channels, the teleportation task is still carried out with the successful probability of $4|df|^2$. Meanwhile, the assistant Charlie merely plays a role of repeater between the sender and the receiver, and thus the assistant should be irresponsible for the safety of the teleportation task.

\textbf{\emph{Remarks}}: According to the aforementioned analyses, we would like to present that:
\begin{enumerate}
  \item Actually, more physical manipulation should be performed by the assistant Charlie than the sender Alice and the receiver Bob in the first application, and the assistant Charlie plays a major role. However, the assistant Charlie is just regarded as the communication port during the whole probabilistic teleportation between the sender Alice and the receiver Bob in the second application. Thus the roles of the assistant Charlie under different communication conditions are quite distinct.
  \item The novel scheme could improve the security of the total teleportation processes from the sender Alice to the receiver Bob, whether the assistant Charlie knows the partially entangled states or not.
  \item Although the above two applications are aimed at the different communication conditions, respectively, each of the two applications would not be realized successfully without our proposal, therefore the new scheme could enlarge the applied range of probabilistic teleportation.
  \item The total successful probabilities for the total teleportation under different communication conditions are equals to $4|df|^2$, and the successful probabilities would be equal to one when each of the two quantum channels is composed of a maximally entangled state.
\end{enumerate}

\section{Conclusion}

In summary, we would like to underline that previous typical schemes are not feasible unless the receiver completely knows the partially entangled state. To overcome this shortage of the former one, we propose a novel scheme for probabilistically teleporting an unknown quantum state via generalized measurements when only the sender has knowledge of the non-maximally entangled channel. The detailed realization procedures of this scheme are also elaborated. Furthermore, a comparison has been made between our protocol and the former ones in terms of the applied conditions, the successful probability of probabilistic teleportation and so on. Moreover, we present two concrete applications of our scheme for probabilistic teleportation of an unknown quantum state from the sender to the receiver with the aid of the assistant based on the different communication conditions, and the applications suggest that the novel proposal should be used to improve the security of the teleportation processes and extend the applied range of probabilistic teleportation.

\ack{This work is supported by the Program for National Natural Science Foundation of China (Grant Nos. 60974037, 61273202, 61134008 and 11074307).}


\section*{References}
\begin{thebibliography}{10}
\bibitem{Bennett} Bennett C, Brassard G, Grepeau C, Jozsa R, Peres A and Wootters W K 1993 Teleporting an unknown quantum statevia dual classical and Einstein-Podolsky-Rosen channels \emph{Phys. Rev. Lett.} 70 1895.
\bibitem{Bouwmeester} Bouwmeester D, Pan J W, Mattle K, Eibl M, Weinfuter H and Zeilinger A 1997 Experimental quantum teleportation \emph{Nature} 390, 575
\bibitem{Boschi} Boschi D, Branca S, Martini F D, Hardy L and Popescu S 1998 Experimental realization of Teleporting an unknown quantum statevia dual classical and Einstein-Podolsky-Rosen channels \emph{Phys. Rev. Lett.} 80, 1121
\bibitem{Karlsson} Karlsson A and Bourennane M 1998 Quantum teleportation using three-particle entanglement \emph{Phys. Rev. A} 58, 4394
\bibitem{Li} Li W L, Li C F and Guo G C 2000 Probabilistic teleportation and entanglement matching \emph{Phys. Rev. A} 61, 034301
\bibitem{Koniorczyk} Koniorczyk M, Kiss T and Janszky J 2001 Teleportation: from probability distributions to quantum states \emph{J. Phys. A: Math. Gen.} 34, 6949-6955
\bibitem{Werner} Werner R F 2001 All teleportation and dense coding schemes \emph{J. Phys. A: Math. Gen.} 34, 7081-7094
\bibitem{Dai1} Dai H Y, Chen P X and Li C Z 2004 Probabilistic teleportation of an arbitrary two-particle state by two partial three-particle entangled W states \emph{J. Opt. B} 6, 106
\bibitem{Dai2} Dai H Y, Chen P X and Li C Z 2004 Probabilistic teleportation of an arbitrary two-particle state by a partially entangled three-particle GHZ state and W state \emph{Opt. Commun.} 231, 281
\bibitem{Yeo} Yeo Y, Liu T Q, Lu Y E and Yang Q Z 2005 Quantum teleportation via a two-qubit Heisenberg XY chain¡ªeffects of anisotropy and magnetic field \emph{J. Phys. A: Math. Gen.} 38, 3235-3243
\bibitem{Zhang2} Zhang Y 2006 Teleportation, braid group and Temperley-Lieb algebra \emph{J. Phys. A: Math. Gen.} 39, 11599-11622
\bibitem{Dai4} Dai H Y, Zhang M and Li C Z 2008 Teleportation of three-level multi-partite entangled state by a partial three-level bipartite entangled state \emph{Commun. Theor. Phys.} 49, 891
\bibitem{Tanaka} Tanaka Y, Asano M and Ohya M 2010 Physical realization of quantum teleportation for a nonmaximal entangled state \emph{Phys. Rev. A} 82, 022308
\bibitem{Neves} Sol\'{\i}s-Prosser M A and Neves L 2011 Remote state preparation of spatial qubits \emph{Phys. Rev. A} 84, 012330
\bibitem{Zhang6} Zhang H L, Liang W D, Liu K, Zhang J X and Gao J R 2012 Fidelity with quadrature component variances for continuous-variable quantum teleportation \emph{J. Phys. B: At. Mol. Opt. Phys.} 45, 115501
\bibitem{Chiribella} Chiribella G, Giovannetti V, Maccone L and Perinotti P 2012 Teleportation transfers only speakable quantum information \emph{Phys. Rev. A} 86, 010304(R)
\bibitem{Nielsen} Nielsen M A and Chuang I L 2000 \emph{Quantum computation and quantum information} (Cambridge: Cambridge University Press)
\bibitem{Dai3} Dai H Y, Chen P X, Liang L M and Li C Z 2006 Classical communication cost and remote preparation of the four-particle GHZ class state \emph{Phys. Lett. A} 355, 285
\bibitem{Peng} Peng Z H, Zou J and Liu X J 2008 Scheme for implementing efficient quantum information processing with multiqubit W-class states in cavity QED \emph{J. Phys. B: At. Mol. Opt. Phys.} 41, 065505
\bibitem{Ikram} Al-Amri M, Evers J, Ikram M. and Zubairy M S 2012 Quantum teleportation of high-dimensional atomic ensemble states \emph{J. Phys. B: At. Mol. Opt. Phys.} 45, 095502
\bibitem{Evers} Al-Amri M, Evers J and Zubairy M S 2010 Quantum teleportation of four-dimensional qudits \emph{Phys. Rev. A} 82, 022329
\bibitem{Zhang} Zhang M, Dai H Y , Zhu X C, Li X W and Hu D W 2006 Control of the quantum open system via quantum generalized measurement \emph{Phys. Rev. A} 73, 032101
\end{thebibliography}
\end{document}